\documentclass[10pt,reprint,twocolumn,aps,prl]{revtex4}
\usepackage{graphicx}
\usepackage{color,soul}
\usepackage{float}
\usepackage{amsmath}
\usepackage{bm}
\usepackage{amssymb}
\usepackage{amsfonts}
\usepackage{dcolumn}
\usepackage{epsfig}
\usepackage{subfigure}
\usepackage{bm}
\begin{document}

\title{Enhanced spin-orbit interaction and Kondo scattering in $\delta$-doped LaTiO$_3$/SrTiO$_3$ interfaces}

\author{Shubhankar Das$^{1}$, A. Rastogi$^{1}$, Lijun Wu$^{2}$, Jin-Cheng Zheng$^{3}$, Z. Hossain$^{1}$, Yimei Zhu$^{2}$ and R. C. Budhani$^{1,4,}$}
\email{rcb@iitk.ac.in, rcb@nplindia.org}
\affiliation{$^1$Condensed Matter - Low Dimensional Systems
Laboratory, Department of Physics, Indian Institute of Technology,
Kanpur 208016, India\\
$^2$Condensed Matter Physics and Material Science department, Brookhaven National Laboratory, Upton, NY 11973, USA\\
$^3$Department of Physics and Fujian Provincial Key Laboratory of Theoretical and Computational Chemistry, Xiamen University, Xiamen 361005, China\\
$^4$National Physical Laboratory, Council of Scientific and Industrial Research (CSIR), New Delhi - 110012, India}

\date{\today}

\begin{abstract}
We present a study of delta ($\delta$) doping at LaTiO$_3$/SrTiO$_3$ (LTO/STO) interface with iso-structural antiferromagnetic perovskite LaCrO$_3$ (LCO) that dramatically alters the properties of the two dimensional electron gas (2-DEG) at the interface. The effects include a reduction in sheet-carrier density, prominence of the low temperature resistivity minimum, enhancement of weak antilocalization below 10 K and observation of a strong anisotropic magnetoresistance (MR). The positive and negative MR for out-of-plane and in-plane field respectively and the field and temperature dependencies of MR suggest Kondo scattering by localized Ti$^{3+}$ moments renormalized by spin-orbit interaction at T $<$ 10 K, with the increased $\delta$-layer thickness. Electron energy loss spectroscopy and density functional calculations provide convincing evidence for blocking of electron transfer from LTO to STO by the $\delta$-layer.
\end{abstract}

\maketitle
The phenomenon of the formation of a 2-dimensional electron gas (2-DEG) at the interface of epitaxially grown LaTiO$_3$ (LTO) or LaAlO$_3$ (LAO) on TiO$_2$ terminated SrTiO$_3$ (STO) \cite{Ohtomo1, Rastogi, Ohtomo2} has attracted much attention in recent years \cite{Siemons, Dikin, Li, Brinkman, Herranz, Biscaras}. It is generally agreed that the gas is formed by transfer of electrons from the polar layer of LAO or LTO to the top TiO$_2$ layer of STO. Since the carrier concentrations (n$_\Box$) are large ($\sim 3\times 10^{14}$/cm$^2$), and some of the Ti$^{4+}$ ions at the interface may also get converted to Ti$^{3+}$ with $S = {\raise0.7ex\hbox{$1$} \!\mathord{\left/  {\vphantom {1 2}}\right.\kern-\nulldelimiterspace} \!\lower0.7ex\hbox{$2$}}$ localized spin, the electron dynamics is likely to be controlled by weak electron-electron (e-e) scattering and magnetic scattering, in addition to the effects of weak static disorder. Moreover, as the interface breaks inversion symmetry, there is a possibility of Rashba spin-orbit scattering \cite{Caviglia} emanating from the interface electric field. Some of these issues have been addressed by measuring the magnetoresistance (MR) of 2-DEG formed at LAO/STO \cite{Franklin, Wang, Joshua} and electrolyte gated STO \cite{Gordon}. However, no consensus has emerged on the origin of a strong positive MR observed when the external magnetic field is perpendicular to the plane of the film (H$_{\bot}$), a change in the sign of the MR when the field is brought in the plane (H$_{\parallel}$), a characteristic minimum in R(T) below $\sim$ 100 K followed by a ln T behavior, and finally, saturation of R(T) at still lower temperatures.

In order to address the mechanism of 2-DEG formation at LTO/STO interface, and to identify the dominant scattering processes that control the nature of MR in this system, we have used a novel approach of $\delta$-doping of the interface. The doped structure consists of LTO(m unit cell(uc))/LCO($\delta$ uc)/TiO$_2$ terminated STO. The LCO/STO alone does not form a 2D gas. The LCO film remains an antiferromagnetic insulator with Cr site spin of ${\raise0.7ex\hbox{$3$} \!\mathord{\left/
{\vphantom {3 2}}\right.\kern-\nulldelimiterspace}\!\lower0.7ex\hbox{$2$}}$ and T$_N$ = 298 K. This is interesting in itself because Cr follows vanadium in the 3d transition series and LaVO$_3$/SrTiO$_3$ interface is conducting \cite{Hotta}. However, when LCO is inserted as a $\delta$-layer, the 2-DEG nature of LTO/STO is retained for smaller values of $\delta$ ($<$ 3), but with the increasing $\delta$, a significant blocking of carriers by LCO drives the interface insulating. The temperature, magnetic field and angular dependence of MR in $\delta$ = 0 indicates a dominant Kondo-type s-d scattering for H$_\parallel$ field. However, the characteristics negative MR of Kondo is superceded by positive MR resulting presumably from the enhanced forward scattering of diffusive electrons by the S-O interaction in the T $\leqslant$ 10 K regime. For H$_{\perp}$, the classical positive MR quadratic in field is seen at T $>$ 10 K. It is interesting to note that the Rashba coupling at the interface of LTO/STO can be modulated by insertion of LCO layers.

The films are deposited using pulsed laser ablation on STO, as described in our earlier works \cite{Rastogi, Rastogi2}. We have deposited three sets of films. In the first set 0, 0.5, 3, 5 and 10 uc of LCO was grown first on STO followed by 20 uc thick LTO film. In the second set the $\delta$ is 5 uc and the LTO was varied from 4 to 24 uc. In the last set, the LTO is 16 uc while LCO is reduced from 5 to 0 uc in steps of 1 uc. The atomic and chemical states of the interface have been studied using X-ray reflectivity and cross sectional scanning transmission electron microscopy (STEM) in conjunction with electron energy loss spectroscopy (EELS). In addition, Density Functional Theory (DFT) calculations have been performed to analyze charge density profile of the interface. Electron transport measurements have been performed in a 14 Tesla ($\mathcal{T}$) system (Quantum Design PPMS) fitted with a sample rotator which allowed measurement of angular MR.

\begin{figure}[h]
\begin{center}
\includegraphics [trim=0cm 0cm 0cm 0cm, angle = 0, width=8.7cm,angle=0]{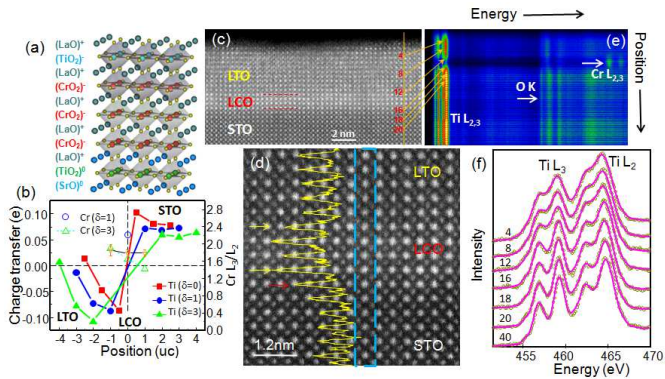}%
\end{center}
\renewcommand{\thefigure}{\arabic{figure}}
\caption{\label{Fig 1} (a) Oxides planes along [001] direction of the $\delta$-doped interface. (b) DFT Charge transfer in (LTO)$_3$(LCO)$_{\delta}$(STO)$_3$. The positive (negative) value means the gain (loss) of charge. The left, middle and right regions are 3 uc of LTO, $\delta$-uc of LCO and 3 uc of STO, respectively. This figure also shows Cr L$_3$/L$_2$ intensity ratio across the interface (brown inverted triangles). (c),(d) HAADF image showing interfaces between LTO, LCO and STO with 3 uc LCO (bright atom columns). An intensity line-profile (yellow) from the column marked by the blue dark line is included in (d). (e) EELS spectrum image from vertical scan line in (c), showing Ti L$_{2,3}$, O K and Cr L$_{2,3}$ edges at $\approx$ 460 eV, 530 eV and 580 eV respectively. (f) a series of Ti L$_{2,3}$ edges (black circles) across two interfaces from the spectrum image (e) acquired from the line scan partially shown in (c) (see text for details).}
\end{figure}

Fig. 1 shows a sketch of various atomic planes of the heterostructure along with high angle angular dark field (HAADF) images taken from STEM. The atomically sharp interfaces and uniformly distributed 3 uc LCO between LTO and STO is clearly seen with bright background contrast due to the high atomic number Z in the LCO unit-cell. The higher peak intensity marked by the red arrows in Fig. 1(d) than the average Sr peak in STO indicates diffusion of La/Cr into STO, limited to 1 to 2 uc. A 2D elemental map based on EELS spectrum image shown in Fig. S1 \cite{Supplementary} also confirms the coherent and atomic sharp interfaces. An EELS image with the Ti L$_{2,3}$, O K and Cr L$_{2,3}$ edges from the vertical scan line in (c) but extended into STO is depicted in Fig. 1(e). The EELS spectra (open dots) as a function of atomic position (Fig. 1(c)) are plotted in Fig. 1(f). The overlaid red lines are results from the multiple linear least square fitting, the spectrum with weighted linear combination of Ti$^{3+}$ and Ti$^{4+}$ reference spectra. Four distinct peaks representing e$_g$ and t$_{2g}$ electron orbital of Ti - L$_2$ and L$_3$ energy level are clearly visible on the STO side and they became broader with peak separation of e$_g$ and t$_{2g}$ less pronounced at the interface and into the LTO side, indicating an increase of the Ti$^{3+}$ state. Composition mapping revealed a constant distribution of oxygen across the region and complementary increase and decrease in Cr and Ti, respectively, in the LCO layer with a 1-2 uc diffusion length \cite{Supplementary}. Since it is known that the Cr$^{2+}$ containing compounds have higher L$_3$/L$_2$ Cr-absorption edge intensity ratio compared to the Cr$^{3+}$ containing compounds \cite{Daulton, Colby}, we have analyzed the L$_3$ and L$_2$ intensities for $\delta$ = 1, 2 and 3 uc samples (see Fig S7 of supplementary section). Our analysis indicates that L$_3$/L$_2$ changes from 1.84 to 1.77 on moving from LTO/LCO interface to LCO/STO interface in the $\delta$ = 3 uc sample. This result suggests that the $\delta$-layer gains electrons from the LTO layer. The percentage of Ti$^{3+}$ over the sum of Ti$^{3+}$ and Ti$^{4+}$ across the interface suggests a significant charge transfer from LTO to STO near the interface. To confirm these findings, we conducted DFT calculations by constructing a supercell with 3 uc LTO on the left, $\delta$ uc LCO in the middle and 3 uc STO on the right \cite{Supplementary}. The calculations show significant charge transfer from LTO to STO (Fig. 1(b)), which reduces with the increase of $\delta$. Interestingly, Cr in LCO also receives electrons, confirming its reduced valence state as suggested by EELS measurements.

\begin{figure}[h]
\begin{center}
\abovecaptionskip -10cm
\includegraphics [width=8.7cm]{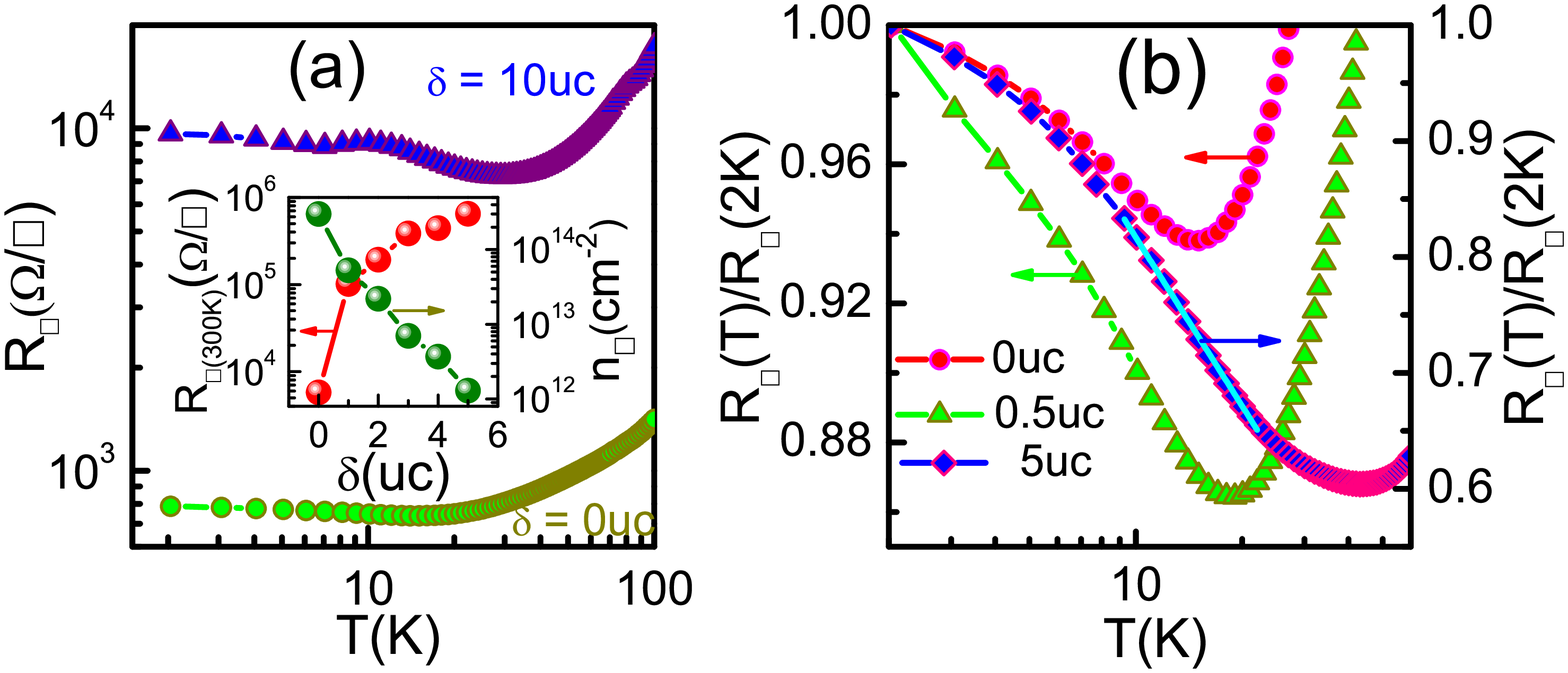}%
\end{center}
\renewcommand{\thefigure}{\arabic{figure}}
\caption{\label{fig2} (a) Temperature dependence of R$_\Box$ of LTO(20 uc)/LCO($\delta$ uc)/STO heterostructure. Inset shows the variation of R$_\Box$ and n$_\Box$ with doping in LTO(16 uc)/LCO($\delta$ uc)/STO. (b) R$_\Box$(T)/R$_{\Box}$(2K) of $\delta$ = 0, 0.5 and 5 uc. The solid line in $\delta$ = 5 uc curve is the ln T fit.}
\end{figure}

Fig. 2(a) shows the sheet resistance (R$_\Box$) as a function of temperature (T) for LTO(20 uc)/LCO($\delta$ uc)/STO samples of $\delta$ = 0 and 10. We see a metallic behavior upon lowering the T from 300 K. On cooling below $\approx$ 20 K, a resistance minimum followed by a slight upturn and then saturation of R$_\Box$ at T $\leqslant$ 7 K is seen for the $\delta$ = 0. As the $\delta$-layer becomes thicker, minimum (T$_m$) shifts towards higher temperature and the upturn becomes more prominent. This trend of R$_\Box$ has been seen in all samples of $\delta$ = 0.5, 3, 5 and 10 uc. Inset of Fig. 2(a) shows R$_\Box$ and n$_\Box$ at 300 K as a function of $\delta$-layer thickness. While R$_\Box$ increases progressively, the n$_\Box$ drops with the increase in $\delta$ layers. For $\delta$ = 0, the n$_\Box$ at 300 K is $\approx$ 3 $\times$ 10$^{14}$ cm$^{-2}$, which is very close to the areal charge density (3.2 $\times$ 10$^{14}$ cm$^{-2}$) expected if half an electron per unit cell is transferred to STO surface from the LTO layers to suppress the polarization catastrophe. The insertion of a few unit cells of LCO leads to a dramatic decrease in n$_\Box$, in fact by a factor of 50 and 280 for $\delta$ = 3 and $\delta$ = 5 uc respectively at 300 K. These observations are consistent with STEM results, which suggest conversion of Cr$^{3+}$ to Cr$^{2+}$ in the LCO layers, and the results of the DFT calculations.

Fig. 2(b) is a plot of the R$_\Box$(T)/R$_\Box$(2 K) of $\delta$ = 0, 0.5 and 5 uc to emphasize the minimum in R$_\Box$(T) at T$_m$. Below T$_m$ the resistance follows a ln T dependence, but this divergence is cutoff on further lowering the temperature. This saturating tendency of R$_\Box$ is prominent in the $\delta$ = 0. The simplest interpretation for the ln T rise can be given in terms of weak localization (WL) in 2D where a constructive interference between partial waves of diffusive electrons can lead to enhance backscattering and hence an increase in resistance, which continues to grow at lower temperatures as the dephasing inelastic scattering is reduced due to phonon freeze out \cite{Bergmann, Lee}. Since weak localization is an orbital effect, it has a distinct dependence on the angle between the H and the plane of the film. H$_\perp$ quenches quantum backscattering because of the Aharonov-Bohm phase acquired by the partial waves. A similar dependence of R$_\Box$ in zero field also results from e-e interaction in 2D \cite{Shao, Liu}. The distinction between the two can be made by measuring the MR, which in the latter case is positive and mostly isotropic. However, before we dwell upon the MR data, a key observation of Fig. 2(b) is the truncation of the divergence of R$_\Box$ at T $\ll$ T$_m$. Such an effect can arise due to a phenomenon closely associative with WL in the presence of the S-O interaction. The dephasing of the spin degree of freedom by S-O in diffusive trajectories can suppress the quantum backscattering and thereby truncate the ln T growth of R$_\Box$ at low temperatures. This weak antilocalization (WAL) \cite{Bergmann} becomes prominent at T $\ll$ T$_m$ as the S-O gains strength at lower temperatures.

Here it is pertinent to introduce one more scattering phenomenon which can lead to a minimum followed by saturation of R$_\Box$ in disordered metallic films. This is the Kondo scattering of conduction electrons of spin $\overrightarrow {S_e }$ by localized magnetic impurity in the system of spin $\overrightarrow {S_i }$. The interaction between the two moments is given by the hamiltonian, $H_{ex}  = J.\overrightarrow {S_i .} \overrightarrow {S_e }$ where J is positive and hence a stable configuration demands anti parallel arrangement of $\overrightarrow {S_i }$ and $\overrightarrow {S_e }$. The Kondo interaction leads to a resistivity $\Delta \rho _k  =  - B\ln T$, here B is a positive constant and a function of J, N(E$_F$) (the density of states at Fermi level) and other properties of the electron gas. However, $\Delta \rho _k$ cannot increase without a bound \cite{Kondo}. Eventually, the divergence of $\Delta \rho _k$ is cutoff and it becomes constant below a temperature of the order of Kondo temperature, $T_K  = T_F \exp ( - {\raise0.7ex\hbox{$1$} \!\mathord{\left/ {\vphantom {1 {J_N }}}\right.\kern-\nulldelimiterspace} \!\lower0.7ex\hbox{${JN }$}})$. This unitary limit is however not reached in metal films \cite{Chandrasekhar, Giordano, Gang}. A H-field suppresses Kondo scattering thereby leading to a negative isotropic MR. Recently, Kondo mechanism has been proposed for R$_\Box$(T, H) of a 2-DEG formed on the surface of STO by electrostatic gating \cite{Gordon}. It has been argued that highly localized 3d$^1$ electrons of some Ti$^{3+}$ ions (spin 1/2) are the source of Kondo scattering. The idea of magnetic scattering is supported by the recent observation of ferromagnetism at LAO/STO interface \cite{Brinkman}.

\begin{figure}[h]
\begin{center}
\includegraphics [width=8.7cm]{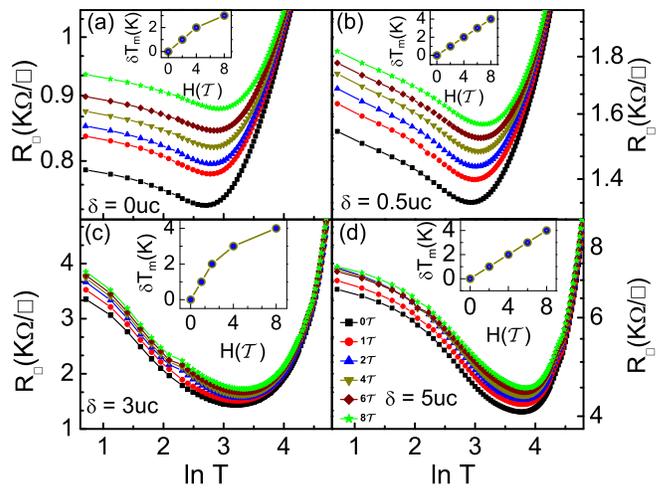}%
\end{center}
\renewcommand{\thefigure}{\arabic{figure}}
\caption{\label{fig3} (a-d) R$_\Box$(T) of LTO(20 uc)/LCO($\delta$ uc)/STO films as a function of ln T for different H$_\perp$. Inset shows $\delta$T$_m$ vs H$_\perp$ plot; where $\delta$T$_m$ = T$_m$(H) - T$_m$(0). All the samples show positive MR down to 2 K.}
\end{figure}

In Fig. 3 we show R$_\Box$(T) at different H$_\perp$ for $\delta$ = 0, 0.5, 3 and 5 uc. The H$_\perp$ shifts resistivity minimum to higher T (see insets) and a dramatic positive MR is evident which is inconsistent with the WL but agrees broadly with the e-e scattering scenario. In the latter case the magnetoconductance goes as                   $ \sim  - \frac{{e^2 }}
{\hbar }\frac{{\tilde F_\sigma  }}
{{4\pi ^2 }}(0.084) \times \left( {\frac{{g\mu _B H}}
{{k_B T}}} \right)^2$ for $\frac{{g\mu _B H}}
{{k_B T}} \ll 1 $, where $\tilde F_\sigma$ has the upper bound of 4/3. Clearly, a positive MR is expected which goes as H$^2$. At high field a ln(H) dependence of MR has been predicted.

\begin{figure}[h]
\begin{center}
\includegraphics [width=8.7cm]{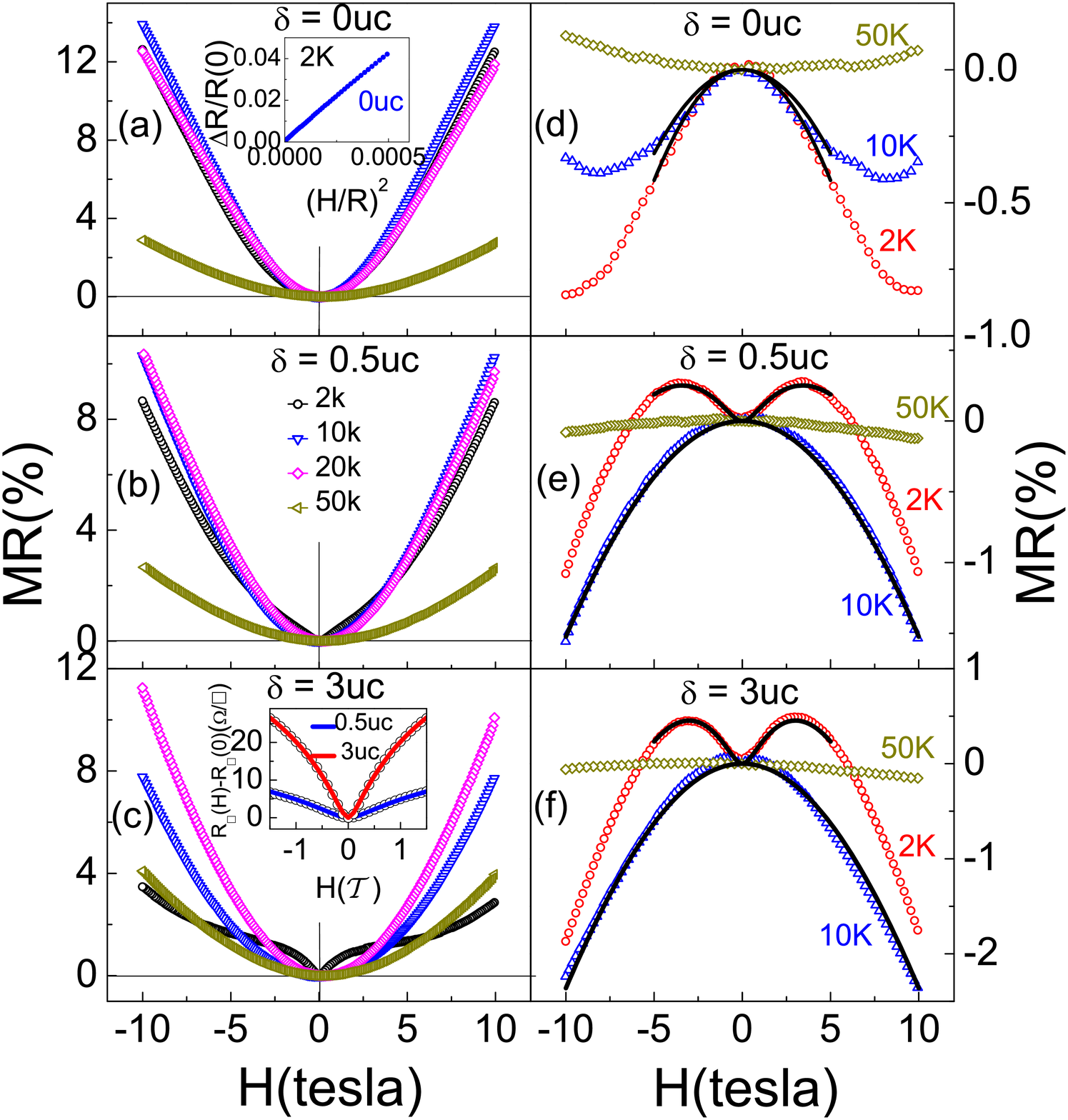}%
\end{center}
\renewcommand{\thefigure}{\arabic{figure}}
\caption{\label{fig4} (a-c) MR$_\perp$ of $\delta$ = 0, 0.5 and 3 uc respectively. Inset of (a) shows Kohler plot for $\delta$ = 0 uc while inset of (c) reveals the WAL effect after subtracting high field $H^2$ data. The solid curves in the inset of (c) are the fit to the Eq. (1). (d-f) show MR$_\parallel$ for the same set of samples. A negative MR$_\parallel$ for all three temperatures is seen for $\delta$ = 0 but the $\delta$ = 0.5 and 3 uc show positive MR$_\parallel$ at lower field at 2 K and a crossover from positive to negative MR at higher field. In Fig. (e, f) the black solid line for 10 K MR$_\parallel$ is the fit using Kondo model (Eq. (2)) and at 2 K it's fitted using Kondo + WAL in the range -5 $\mathcal{T}$ $\leqslant$ H $\leqslant$ 5 $\mathcal{T}$.}\label{fig4}
\end{figure}

We probe the MR further as a function of H-field. A Positive ($\approx$14\%) out-of-plane MR (MR$_\perp$) for $\delta = 0$ uc is observed at 2 K and 10 $\mathcal{T}$ (fig. 4(a)). The MR$_\perp$ has a H$^2$ dependence, which, at first glance can be attributed to the e-e scattering.  The upper bound for H to see H$^2$ dependance at 4.2 K is $\approx$ 3.16 $\mathcal{T}$ and the slope of MR vs H$^2$ curve is $\approx$ 0.714 $\times$ 10$^{-7}$ /$\mathcal{T}$$^2$ (calculated from the e-e scattering theory). However, the measured slope for $\delta$ = 0 is 1.69 $\times$ 10$^{-3}$ /$\mathcal{T}$$^2$, which suggests that the e-e interaction alone is not responsible for the large MR$_\perp$. A sizable contribution to MR$_\perp$ can also come from the classical defect scattering \cite{Tufte} that follows the Kohler's rule: $\frac{{\Delta R}}{{R_0 }} \propto a[\frac{H}{{R_0 }}]^2$. The inset of Fig. 4(a) shows Kohler's plot for the $\delta$ = 0. From these MR$_\perp$ data the mobility of carriers at 2 K and 100 K comes out to be 403 and 86 cm$^2$-V$^{-1}$-S$^{-1}$ respectively.

Fig. 4(b) and 4(c) show that the MR$_\perp$ at 2 K and 10 $\mathcal{T}$ for $\delta$ = 0.5 and 3 uc decreases to 9\% and 4\% respectively. At lower fields it also deviates from H$^2$ and a cusp appear near H = 0. This indicating the presence of an additional scattering mechanism that becomes operational below $\approx$ 10 K. We separate out the contribution of this process by extrapolating the H$^2$ dependence seen at H $\geqslant$ 6 $\mathcal{T}$ to lower fields and then subtracting the extrapolated value from the measured R$_\Box$(H) (inset of Fig. 4(c) for $\delta$ = 0.5 and 3 uc). We attribute this distinct contribution to MR$_\perp$ at T $\leqslant$ 10 K to S-O scattering, which in the 2D limit for H$_\perp$ can be expressed as \cite{Shao, Hikami, Boris, Lee};
\begin{eqnarray}
\frac{{\Delta R_\square  (H)}}
{{[R_\square  (0)]^2 }} =  - \frac{{e^2 }}
{{2\pi ^2 \hbar }}[\Psi (\frac{1}
{2} + \frac{{H_\varphi  }}
{H}) - \ln \frac{{H_\varphi  }}
{H}]
\end{eqnarray}
Where $\Delta R_\square  (H) = R_\square  (H) - R_\square  (0)$, $\Psi(x)$ is the digamma function and $H_\varphi   = \hbar /(4eL_\varphi ^2 )$. The length $L_\varphi   = \sqrt {D\tau _\varphi  }$ where D and $\tau _\varphi$ are diffusion constant and phase coherence time respectively. Inset of Fig. 4(c) shows the fits of Eq. (1) to MR$_\perp$ of $\delta$ = 0.5 and 3 uc. This yields $L_\varphi$ of $\approx$ 33 nm and 46 nm for $\delta$ = 0.5 and 3 uc respectively. These numbers are reasonable considering the fact that the scattering is taking place in the plane of the film where $L_\varphi$ has no dimensional constraints.

Fig. 4(d-f) shows MR$_\parallel$ of $\delta$ =0, 0.5 and 3 uc films. Interestingly, for $\delta$ = 0 we have a negative MR$_\parallel$ at T $<$ 50 K. The suppression of classical positive MR can be understood as the thickness of the 2-DEG is within one  carrier mean free path.  This MR anisotropy also supports the 2D nature of the metallic state in these interfaces.

Interesting MR$_\parallel$ is seen in Fig. 4(e) and 4(f) for $\delta$ = 0.5 and 3 uc respectively at 2 K. Here, the data can be divided in two regions, a positively sloped MR at lower field and a negatively sloped MR at the higher field, resulting in a local MR maximum seen at 3.6 and 3.2 $\mathcal{T}$ for $\delta$ = 0.5 and 3 uc respectively. In our samples this maxima is observed at a much higher field than in 2D metal films of Bi and Au where the crossover field is $\sim$ 0.1 and 2.5 $\mathcal{T}$ respectively \cite{Franklin, Komnik, Kawaguti}. This in-plane positive MR$_\parallel$ diminishes above $\sim$5 K.

The negative MR$_\parallel$ supports the Kondo mechanism. To establish in further, we fit the MR$_\parallel$ of $\delta$  = 0, 0.5 and 3 uc at 10 K to a simple Kondo model \cite{Gordon}
\begin{eqnarray}
R^{model} \left( {H_\parallel  } \right) = R_0  + R_K \left( {H_\parallel  /H_1 } \right)
\end{eqnarray}
where $R_0$ is the residual resistance, $R_K \left( {H_\parallel  /H_1 } \right)$ is a function for zero temperature MR of Kondo impurity, which is related to magnetization and can be calculated using Bethe-ansatz technique \cite{Supplementary}, and $H_1$ is a H-field scale related to T$_K$ and g-factor of impurity spin \cite{Andrei}. The MR$_\parallel$ at 2 K of $\delta$ = 0 uc also fits to the Kondo model (Eq. 2). We note that the negative MR$_\parallel$ at 10 $\mathcal{T}$ (Fig. 4(e, f)) increases with $\delta$-layer thickness and thus bears an inverse relation with n$_\Box$ (see Fig. 2(a)). In Kondo theory R$_K$(T=0, H=0) $\propto$ n$_\Box$$^{-1}$N(E$_F$)$^{-1}$ \cite{Costi}. The data shown in Fig. 4(e, f) are consistent with this picture.

The positive MR$_\parallel$ at 2 K in $\delta$ $\neq$ 0 at fields below a critical value appears to be the contribution of the WAL. To fit the 2 K data we add the WAL and Kondo terms (Eq. (1) and (2)). As the WAL effect is insignificant at higher fields, we fit the 2 K data in the range -5 $\mathcal{T}$ $\leqslant$ H $\leqslant$ 5 $\mathcal{T}$. The black line in Fig. 4(e) and (f) for the 2 K data is this fit \cite{Supplementary}.  The quality of fit strongly suggests that the WAL effect rides over the Kondo scattering at T $<$ 10 K.

The MR of the $\delta$ = 0, 0.5 and 3 uc for different orientation ($\theta$) of the H with respect to sample normal has been measured (see fig. S6 of supplementary). As we tilt the H towards sample plane, a crossover from positive MR to negative MR is observed. This change of sign at 10 $\mathcal{T}$ happens at $80^0$, $70^0$ and $50^0$ for $\delta$ =0, 0.5 and 3 uc respectively. The angular variation of R$_\Box$ is of the type $R(\theta ,T) = R(T)\cos ^2 (\theta ) + R_0 (T)$, where R(T = 2 K) = 33, 36, 44 $\Omega$ and R$_0$(T = 2 K) = 233, 466, 906 $\Omega$ for $\delta$ = 0, 0.5, 3 uc respectively.

In summary, We have established a strong suppression of n$_\Box$ in 2-DEG at LTO/STO interface by inserting $\delta$-thick layer of an iso-structural perovskite LCO. Our spectroscopic measurements suggest that Cr ions at the interface act as traps and absorb electron donated by the LTO. The saturation tendency of resistance at T $\leqslant$ 10 K and the ln T dependence between 10 K and T$_m$ are consistent with the Kondo scattering of electrons by localized spins. The origin of the latter can be attributed to electrons in Ti d$^1$ configuration which are presumably in Ti$_{xy}$ orbitals forming heavy polarons with spin S = 1/2, while the conduction takes place in extended band of Ti$_{yz/zx}$ motif \cite{Gordon, Salluzzo, Kim, Nanda}. Such Ti$^{3+}$ site will presumably have zero spin due to complete delocalization of 3d$^1$ electron. We also argue that the interfacial Cr$^{3+}$ ions (S = 3/2) may also contribute to s-d scattering. However, as most of the Cr$^{3+}$ spins are antiferromagnetically ordered, such a contribution may come only from the disordered spins located at the LaCrO$_3$-SrTiO$3$ interface. Our STEM results shown in Fig. 1(d) do indicate some diffusion of La/Cr into STO. Further, if some of the Cr$^{3+}$ ions are converted into Cr$^{2+}$ as indicated by our EELS measurements and also suggested by the depletion of 2DEG carrier density on $\delta$-doping, the site spin of Cr$^{3+}$ would deviate from S= 3/2 and affect the antiferromagnetic arrangement. The emergence of a cusp in the positive MR for H$_{\bot}$ in $\delta$-doped samples at T $<$ 10 K is in agreement with the prediction of 2D-WAL theory as evidence by the large value of $L_{\varphi}$. The 2D-WAL also couples with Kondo MR response of the sample at T $<$ 10 K and H$_{\parallel}$ $\leqslant$ 3 $\mathcal{T}$. An important finding of this work is the enhanced S-O interaction in the presence of $\delta$-layer. In the Rashba scenario, it needs to be seen how the $\delta$-layer enhances the local electric field at the interface.

We thank P. C. Joshi for technical support. RCB acknowledges J. C. Bose National Fellowship of the Department of Science and technology, Government of India. SD and AR thank IIT Kanpur and CSIR for financial support. This research has been funded by the CSIR-India and IIT Kanpur. Work at BNL was supported by U. S. Department of Energy, office of Basic Energy Science, under contract No. DE-AC02-98CH10886. JCZ acknowledge the support of NSF of China (grand no. U1232110).

\end{document}